\documentclass[owncolumn,showpacs,preprintnumbers,amsmath,amssymb]{revtex4}
\usepackage{graphicx}% Include figure files
\usepackage{dcolumn}% Align table columns on decimal point
\usepackage{bm}% bold math

\begin{document}

\title{Cosmological Acceleration from Virtual Gravitons}% Force line breaks with \\

\author{Leonid Marochnik}
\email{lmarochnik@gmail.com}
 \affiliation{Physics Department, University of Maryland, College Park, MD 20742, USA}
\author{Daniel Usikov}%
 \email{dusikov@electroglas.com}
\affiliation{36477 Buckeye St., Newark, CA 94560, USA}
\author{Grigory Vereshkov}
\email{gveresh@gmail.com} \affiliation{Research Institute of
Physics, Southern Federal University, 344090, Rostov--on--Don,
Russia}

\begin{abstract}

Intrinsic properties of the space itself and quantum fluctuations
of its geometry are sufficient to provide a mechanism for the
acceleration of cosmological expansion (dark energy effect).
Applying Bogoliubov--Born--Green--Kirkwood--Yvon hierarchy
approach to self--consistent equations of one--loop quantum
gravity, we found exact solutions that yield acceleration. The
permanent creation and annihilation of virtual gravitons is not in
exact balance because of the expansion of the Universe. The excess
energy comes from the spontaneous process of graviton creation and
is trapped by the background. It provides the macroscopic quantum
effect of cosmic acceleration.
\end{abstract}

\pacs{Dark energy 95.36.+x; Quantum gravity 04.60.-m}

\maketitle

\section{Introduction}

To explain the dark energy effect \cite{1, 2} a cosmological
constant, hypothetical fields or some modifications of physical
laws have been proposed (see review \cite{3}). We show that
quantum fluctuations of the metric (gravitons) and their back
reaction on the isotropic and homogeneous (on average) background
provide the mechanism for cosmological acceleration. The  dark
energy effect is a consequence of the vacuum polarization and
graviton creation by non--stationary gravitational field of the
Universe.

The energy density of gravitons is a functional of the background
geometry. In the non--empty Universe the background geometry is
defined by all contributing cosmological subsystems --- by
gravitons, matter and radiation. The combination of conformal
non--invariance with zero rest mass of gravitons (unique
properties of the gravitational field) leads to a macroscopic
quantum effect: condensation of gravitons in a quantum state with
wavelength of the order of the distance to the horizon. In the
process of the evolution of the Universe, the density of
gravitons, as a result of their condensation, starts dominating
over the sum total of the energy density of other subsystems of
the cosmological media. The self--consistent state of background
and gravitons, which evolves asymptotically, represents
self--polarized vacuum in the de Sitter space. In the subsequent
paper \cite{4} we show that the regime of the de Sitter-like
expansion is beginning to form in the current Universe which is
consistent with dark energy observations  \cite{5}.

We present three new exact solutions for the one--loop quantum
gravity. Two of these provide cosmological expansion with
acceleration. The de Sitter solution is one of these. All exact
solutions can be found when the theory is presented as
Bogoliubov--Born--Green--Kirkwood--Yvon (BBGKY) hierarchy
equations for moments of the graviton spectral function. The same
results follow from direct calculations of quantum field operator
functions, state vectors and the graviton spectral function.

We operate in the framework of one--loop quantum gravity because
the theory cannot be renormalized in higher loops. Problems
arising in two--loop theory are described, for example, in
\cite{6}. However, the effect of condensation of gravitons is
created by general properties of gravitational field (conformal
non--invariance and zero rest mass of gravitons). These properties
will probably remain in the future comprehensive theory which
includes quantum gravitation. The results of one--loop theory in
which only the gravitational field is taken into account, are
mathematically robust due to the finiteness of one--loop quantum
gravitation \cite{7}. In our case, the finiteness is provided by
the compensation of diverged contributions of gravitons and ghosts
to observable quantities.

\section{Equations}

The complete classic (non-quantum) theory of back reaction of
scalar, vector and tensor fluctuations on the isotropic
space--time background has been developed in \cite{8}. In this
paper, we consider the quantum theory of tensor fluctuations. Our
model of the empty Universe consists of the background and
gravitons only. In the self--consistent theory of gravitons, the
macroscopic metric is described by regular Einstein equations
\begin{equation}
 \displaystyle R_i^k-\frac12\delta_i^kR=\varkappa
 \langle\Psi|\hat T_{i(grav)}^k+\hat T_{i(ghost)}^k|\Psi\rangle\ .
 \label{1}
 \end{equation}

The stress tensor of gravitons $\hat T_{i(grav)}^k$ and ghosts
$\hat T_{i(ghost)}^k$ should be obtained by solving operator
equations of motion and averaging over a quantum ensemble
$|\Psi\rangle$. Note the average stress tensor of nontrivial ghost
fields interacting with gravity must appear in the right hand side
of (\ref{1}) because there are no gauges that eliminate the
diffeomorphism group degeneracy in the General Relativity. Our
gauge selection was based on two principles. First, both
background and gravitons should be considered in the same
reference frame. Second, the gauge should provide automatically
the one--loop finiteness. We found the only gauge that satisfies
both conditions comes from the set of synchronic gauges.

Our calculations were done in the frame of one--loop approximation
over quantum fields. In the flat isotropic Universe, the 
equations (\ref{1}) read
\begin{equation}
\displaystyle
3H^2=\varkappa\varepsilon_g\equiv\frac{1}{16}D+\frac14W_1\ ,
\qquad -2\dot H-3H^2=\varkappa
p_g\equiv\frac{1}{16}D+\frac{1}{12}W_1\ , \label{2}
\end{equation}
where $H=\dot a/a$ is the Hubble function and $a(t)$ is the scale
factor. Here $D$ and $W_1$ are moments of the spectral
distribution function of gravitons that is renormalized by ghosts.
The moments are:
\begin{equation}
\begin{array}{c}
\displaystyle D=\ddot W_0+3H\dot W_0\ ,
\\[3mm]
\displaystyle W_m=\sum_{{\bf
k}}\frac{k^{2m}}{a^{2m}}\left(\sum_\sigma\langle g|\psi^+_{{\bf
k}\sigma}\psi_{{\bf k}\sigma}|g\rangle-\langle gh|\bar\theta_{{\bf
k}}\theta_{{\bf k}}|gh \rangle\right),\qquad
 m=0,\,1,\,2,\,...,\infty.
\end{array}
\label{3}
 \end{equation}
Here and later the dots are time derivatives. Heisenberg's
equations for Fourier components of the transverse 3--tensor
graviton field and Grassman ghost field are:
\begin{equation}
\begin{array}{c}
 \displaystyle  \ddot \psi_{{\bf k}\sigma}+3H\dot \psi_{{\bf
k}\sigma}+\frac{k^2}{a^2}\psi_{{\bf k}\sigma}=0\ ,
\\[3mm]
\displaystyle  \ddot \theta_{{\bf k}}+3H\dot \theta_{{\bf
k}}+\frac{k^2}{a^2}\theta_{{\bf k}}=0\ .
\end{array}
\label{4}
 \end{equation}
Taking account of normalization of fields in accordance with
(\ref{2}) and (\ref{3}), canonical commutation relations for
gravitons and anticommutation relations for ghosts read
\begin{equation}
\begin{array}{c}
 \displaystyle \frac{a^3}{4\varkappa}\left[\dot\psi^+_{{\bf k}\sigma},\
\psi_{{\bf k'}\sigma'}\right]_{-}=-i\hbar \delta_{{\bf k}{\bf
k'}}\delta_{\sigma\sigma'}\ ,
\\[3mm] \displaystyle
\frac{a^3}{8\varkappa}\left[\dot{\bar\theta}_{{\bf k}},\
\theta_{{\bf k'}}\right]_+=
-\frac{a^3}{8\varkappa}\left[\dot{\theta}_{{\bf k}},\
\bar\theta_{{\bf k'}}\right]_+=-i\hbar \delta_{{\bf k}{\bf k'}}\ .
\end{array}
\label{5}
 \end{equation}

Equations (\ref{2}, (\ref{3}), (\ref{4}) and quantization rules
(\ref{5}) have been obtained by the path integral \cite{9, 10}.
They have been obtained from the class of synchronic gauges that
automatically provide one--loop finiteness of observables. One--loop
effects of vacuum polarization and particle creation by background
field are contained in equations (\ref{4}) for gravitons and
ghosts. These equations are linear in quantum fields but their
coefficients depend on the non--stationary background metric.
Correspondingly, in the background equation (\ref{2}) we keep the
average values of bilinear forms of quantum fields only. In this
model, quantum particles interact through a common
self--consistent field only.

We would like to point out that our model contains the graviton
energy stress tensor but the short wavelength approximation is not
used. The separation of the total metric into background and
gravitons is not done on the basis of space scales hierarchy  but
on the basis of symmetry criteria. The metric of the isotropic
3--space belongs to 3--scalar representation of the group  $O(3)$
while the gravitons are represented by transverse and traceless
3--tensor. The mathematically rigorous method of separating the
background and the gravitons, which ensures the existence of the
graviton energy stress tensor is based on averaging over graviton
polarizations: $\langle\psi_\alpha^\beta\rangle\equiv 0$ if all
polarizations are equivalent in the quantum ensemble. The
properties of the theoretical model are quite specific, but it is
due to this that we can study the effect of condensation of long
wavelength gravitons in the isotropic universe.

\section{BBGKY hierarchy and exact self--consistent solutions}

The following BBGKY hierarchy has been obtained from (\ref{4}) by
a standard procedure
\begin{equation}
\begin{array}{c}
\displaystyle  \dot D+6HD+4\dot W_1+16HW_1=0\ , \\[3mm]
 \displaystyle
\stackrel{...}{W}_m +3(2m+3)H\ddot{W}_m  +3\left[ \left(
4m^2+12m+6\right)H^2
+(2m+1)\dot{H} \right]\dot{W}_m+ \vspace{5mm} \\
\displaystyle +2m\left[ 2\left(2m^2+9m+9\right)H^3
+6(m+2)H\dot{H}+\ddot{H} \right]W_m
+4\dot{W}_{m+1}+8(m+2)HW_{m+1}=0, \qquad \displaystyle
m=1,\,...,\,\infty\, .
\end{array}
\label{6}
\end{equation}
In the infinite BBGKY chain (\ref{6}), each equation connects two
neighboring moments (\ref{3}) of spectral function.    The system
of equations (\ref{2}), (\ref{6}) has at least three exact
self--consistent solutions. Two of them are the following
\begin{equation}
\begin{array}{c}
\displaystyle  D=
 -48\left[\pm\frac{K^2}{a^2}\left(\ln\frac{a}{a_0}+\frac12\right)-\frac{C}{a^6}\right]\, ,
 \qquad
W_1=\pm 24\frac{K^2}{a^2}\left(\ln\frac{a}{a_0}+\frac14\right)\, ,
\\[3mm]
\displaystyle W_m=-24(\mp
1)^m\frac{K^{2m}}{a^{2m}}\ln\frac{a}{a_0}\, ,\qquad m\geqslant 2\
;\qquad\qquad
  H^2=\pm\frac{K^2}{a^2}\ln\frac{a}{a_0}+\frac{C}{a^6}\ ,
\end{array}
\label{7}
\end{equation}
where $K^2,\, a_0, \, C$ are arbitrary constants. Evolution
scenarios for (\ref{7}) are simplest with $C=0$. The first
solution (upper signs in (\ref{7})) describes the Universe that
was collapsing in the infinitely remote past to the state with the
minimal scale factor $a_{min}=a_0$, and then began to expand with
acceleration $\ddot a/a=K^2/2a^2$. Asymptotically, it becomes
logarithmically slow and reads
\[
\displaystyle a(t)\to Kt\ln^{1/2}(Kt/a_0), \qquad  t\to\infty\ .
\]

The second solution (lower signs in (\ref{7})) corresponds to the
Universe creation from a singularity, expanding to the maximal
scale factor $a_{max}=a_0$, then subsequently collapsing and
ending in a final singularity.

The third solution describes the graviton vacuum in the de Sitter space. It reads
\begin{equation}
\begin{array}{c}
\displaystyle H=\frac16\sqrt{W_1}\ ,\qquad
 a=a_0e^{Ht}\ ,\qquad  D=-\frac{8}{3}W_1\ ,\qquad
  W_{m+1}=-\frac{m(2m^2+9m+9)}{2(m+2)}H^2W_m\ ,\qquad m\geqslant 1\ .
\end{array}
\label{8}
 \end{equation}
One can show that this solution is stable against small perturbations.

From the recurrence relation for moments (\ref{8}), we can
evaluate graviton and ghost characteristic wave lengths  as
follows
\begin{equation}
 \displaystyle \lambda\sim \frac{a}{\overline{k}}\sim \sqrt{
 \frac{W_1}{|W_2|}}=\frac{1}{H}\sqrt{\frac{3}{10}}=const\ .
 \label{9}
 \end{equation}
Quantum fluctuations of these wave lengths dominate in the
formation of observables. As it is seen from (\ref{9}), in the
process of the Universe exponential expanding characteristic
values of $\overline{k}$  rapidly shift to the region of
exponentially large wave numbers, and all observables behave as
constants with time. We will see below that this situation takes
place if the graviton and ghost spectra are flat in ${\bf k}$ ---
space of conformal wave numbers, and divergent contributions from
graviton and ghost integrals compensate each other, so values of
observables are generated by finite differences of these
integrals.

\section{Graviton and ghost state vectors in de Sitter space}

Below we show that the de Sitter solution (\ref{8}) follows from
direct calculations of quantum field operator functions, state
vectors and graviton spectral function, i.e. independently of
BBGKY--approach. Due to the exact solution (\ref{8}), there is a
vacuum state vector such that exact solutions of operator
equations (\ref{4}) become self--consistent in the de Sitter space
averaged over this vector. These exact solutions are
\begin{equation}
\begin{array}{c}
\displaystyle  \psi_{{\bf k}\sigma}=\frac{1}{a}\sqrt{\frac{2\varkappa\hbar}{k}}
 \left[c_{{\bf
 k}\sigma}f(x)+
c^+_{{\bf
 -k}-\sigma}f^*(x)\right]\ ,
 \\[3mm]
 \displaystyle \theta_{{\bf k}}=\frac{1}{a}\sqrt{\frac{4\varkappa\hbar}{k}}
 \left[\alpha_{{\bf
 k}}f(x)+
\bar \beta_{{\bf
 -k}}f^*(x)\right]\ ,
\end{array}
\label{10}
 \end{equation}
where $f(x)=(1-i/x)e^{-ix}$, $x=k\eta$,  $\eta=\int{dt/a}$.
Substituting (\ref{10}) into (\ref{5}), we get
commutation/anticomutaition relations for operator constants:
\begin{equation}
 \displaystyle\left[c_{{\bf  k}\sigma},\ c^+_{{\bf  k'}\sigma'}\right]_-=\delta_{{\bf
 kk'}}\delta_{\sigma\sigma'}\ ,
\label{11}
 \end{equation}
\begin{equation}
 \displaystyle \left[\alpha_{{\bf  k}},\ \bar\alpha_{{\bf  k'}}\right]_+=\delta_{{\bf
 kk'}}\ ,\quad
 \displaystyle \left[\beta_{{\bf  k}},\ \bar\beta_{{\bf
k'}}\right]_+=-\delta_{{\bf
 kk'}}\ .
\label{12}
\end{equation}

In accordance with (\ref{11}), the space of graviton states is
constructed over the standard Fock basis $| n_{{\bf
k}\sigma}\rangle$. Note that quantum occupation numbers  $\
n_{{\bf  k}\sigma}=0,\ 1,\ 2,\ ...,$  cannot be interpreted as
real graviton numbers. In non--stationary space quanta with
arbitrary momentum are not in the mass shell. So, formally, these
are virtual particles, and they behave as real particles
asymptotically only (when $k\eta\gg 1$). Quanta with the
characteristic wave length (\ref{9}) are essentially virtual
particles. Thus, occupation numbers  $n_{{\bf  k}\sigma}$  are
parameters of polarized vacuum. The vacuum state vector is a
product of superposition of states with different occupation
numbers:
\begin{equation}
\displaystyle |g\rangle=\prod_{{\bf  k}\sigma}\sum_{n_{{\bf
k}\sigma}}\mathcal{C}_{n_{{\bf  k}\sigma}}|n_{{\bf
k}\sigma}\rangle\ ,\quad \sum_{n_{{\bf
k}\sigma}}|\mathcal{C}_{n_{{\bf  k}\sigma}}|^2=1\ .
\label{13}
\end{equation}
where $\mathcal{C}_{n_{{\bf  k}\sigma}}$ is the amplitude of the
state with the occupation number $n_{{\bf  k}\sigma}$.

According to (\ref{2}), (\ref{3}), (\ref{4}), the ghosts that
correspond to the set of synchronic gauges redefine the moments of
spectral distribution function additively. In the one--loop
quantum gravity they should provide finiteness values of the
observables. From the physical point of view, ghost fields act as
compensators. We will show later that ghosts provide exact zero
energy of quantum (and quasi--classical) gravitational waves,
which cannot exist in de Sitter space.

Grassman units are extracted from ghost operator constants multiplicatively. They read
\begin{equation}
  \begin{array}{c}
 \displaystyle\alpha_{{\bf  k}}=ua_{{\bf  k}}\ , \qquad
\bar\alpha_{{\bf  k}}=\bar ua^+_{{\bf  k}}\ , \qquad\qquad
\beta_{{\bf k}}=\bar ub_{{\bf  k}}\ , \qquad \bar\beta_{{\bf k}}=
ub^+_{{\bf k}}\ .
\end{array}
 \label{14}
 \end{equation}
By definition, $\bar uu=-u\bar u=1$, so operators $a_{{\bf k}}$,
$a^+_{{\bf  k}}$, $b_{{\bf k}}$, $b^+_{{\bf  k}}$,  which were
introduced in (\ref{14}), satisfy standard Bose commutation
relations. Then, the procedure to construct the ghost space of
states is obvious. The ghost state vector is
\begin{equation}
\begin{array}{c}
\displaystyle |\ gh\rangle=\prod_{{\bf  k}}\sum_{n_{{\bf
k}}}\mathcal{A}_{n_{{\bf  k}}}|\ n_{{\bf k}}\rangle\prod_{{\bf
k}}\sum_{\bar n_{{\bf k}}}\mathcal{B}_{\bar n_{{\bf  k}}}|\ \bar
n_{{\bf k}}\rangle\ , \qquad\qquad \sum_{n_{{\bf
k}}}|\mathcal{A}_{n_{{\bf k}}}|^2= \sum_{\bar n_{{\bf
k}}}|\mathcal{B}_{\bar n_{{\bf k}}}|^2=1\ .
\end{array}
\label{15}
\end{equation}

\section{Calculation of observables}

Further, we make use of the de Sitter solution $a=-(H\eta)^{-1}$
and variable of integration $x=k\eta$. Substituting (\ref{10})
into (\ref{3}) and averaging over state vectors (\ref{13}),
(\ref{15}) one gets
\begin{equation}
  \begin{array}{c}
 \displaystyle
 W_m=\frac{2\varkappa\hbar}{\pi^2}H^{2m+2}\int\limits_0^{\infty}dxx^{2m+1}
 \biggl\{U_{{\bf k}(wave)}|f(x)|^2 +U_{{\bf k}(cr)}[f^*(x)]^2+U_{{\bf k}(ann)}[f(x)]^2\biggr\}
 \ ,
\end{array}
 \label{16}
 \end{equation}
 where
 \[
 \begin{array}{c}
 \displaystyle
 U_{{\bf k}(wave)}=\sum_\sigma\langle g|c^+_{{\bf k}\sigma}c_{{\bf
 k}\sigma}|g\rangle-
  \langle gh|a^+_{{\bf k}}a_{{\bf
 k}}|gh\rangle-\langle gh|b^+_{{\bf k}}b_{{\bf
 k}}|gh\rangle\ ;
\end{array}
\]
\[
 \begin{array}{c}
\displaystyle  U_{{\bf k}(cr)}=\frac12\sum_\sigma\langle g| c^+_{{\bf
k}\sigma}c^+_{{\bf
 -k}-\sigma}|g\rangle-\langle gh| a^+_{{\bf
k}}b^+_{{\bf
 -k}}|gh \rangle\ ;
\\[3mm]
\displaystyle  U_{{\bf k}(ann)}=\frac12\sum_\sigma\langle g| c_{{\bf
-k}-\sigma}c_{{\bf
 k}\sigma}|g\rangle-\langle gh| b_{{\bf
-k}}a_{{\bf
 k}}|gh\rangle\equiv U^*_{{\bf k}(cr)}
\end{array}
 \]
Here $U_{{\bf k}(wave)}$ is the spectral parameter of quantum
waves, which are real gravitons if $k\eta\gg 1$, and $U_{{\bf
k}(cr)}$, $U_{{\bf k}(ann)}$ are spectral parameters of quantum
fluctuations that emerge in processes of graviton creation from
the vacuum and graviton annihilation to the vacuum.

Two conditions for $W_m$, the absence of divergences and absence
of $t$ dependence, should be considered jointly. Because of
$k=x/\eta$ dependence, $U_{{\bf k}(wave)}$ term in (\ref{16}) is
time independent only if $U_{{\bf k}(wave)}$ does not depend on
${\bf k}$. If, however, $U_{{\bf k}(wave)}=const\ ({\bf k})\ne 0$
then this integral does not exist because of the $|f(x)|^2\to 1$
condition at $x\to \infty$. This is the reason why an exact
compensation of graviton and ghost contributions in $U_{{\bf
k}(wave)}$ is a mandatory requirement. The compensation condition
leading to $U_{{\bf k}(wave)}=0$ is
\begin{equation}
 \displaystyle  |\mathcal{C}_{n_{{\bf k}\sigma}}|=|\mathcal{A}_{n_{{\bf
k}}}|=|\mathcal{B}_{n_{-{\bf k}}}|\ .
 \label{17}
 \end{equation}
This result has a simple physical interpretation. Quantum
gravitational waves, whose equation of state is not
$p=-\varepsilon$, cannot exist in de Sitter space with the
self--consistent geometry.

Analogously, spectral parameters $U_{{\bf k}(cr)}=U$, $U_{{\bf
k}(ann)}=U^*$ cannot depend on ${\bf k}$ also.  But corresponding
integrals in (\ref{16}) do not lead to obligatory divergences
because oscillating functions $\sim e^{\pm 2ix}$ are integrated at
$x\to \infty$. Such integrals may be redefined to have certain
finite values. Thus, in (\ref{16}) we have flat graviton and ghost
spectra, $U_{{\bf k}(wave)}\equiv 0$, $U_{{\bf k}(cr)}=U^*_{{\bf
k}(ann)}=U$ and
\begin{equation}
 \begin{array}{c}
 \displaystyle  U=\left(\sum_n\mathcal{C}^*_{n+1}\mathcal{C}_n\sqrt{n+1}\right)^2-
 \left(\sum_n\mathcal{A}^*_{n+1}\mathcal{A}_n\sqrt{n+1}\right)
 \left(\sum_n\mathcal{B}^*_{n+1}\mathcal{B}_n\sqrt{n+1}\right)\ ,
 \\[3mm]
 \displaystyle  |\mathcal{C}_n|=|\mathcal{A}_n|=|\mathcal{B}_n|\equiv \sqrt{\mathcal{P}_n} \ ,
 \end{array}
 \label{18}
 \end{equation}
where $\mathcal{P}_n$ is any universal distribution and $
\displaystyle\bar n=\sum_{n=0}^\infty n\mathcal{P}(n)$ is the
average number of gravitons with wave lengths that are near the
characteristic value (\ref{9}).

From (\ref{17}) and (\ref{18}) it follows that $U=0$ if all
amplitudes are real. We have $U\ne 0$ if some amplitude are
complex and phase dependencies of occupation numbers are different
in graviton and ghost sectors. Observable quantities are
proportional to $U+U^*=2N_g$, where  $N_g$ is a generalized
parameter of graviton vacuum state. To use some model
distributions (Poisson distribution, for example) one can see that
$N_g\sim \bar n$.

Integrals in (\ref{16}) should be redefined. They can be calculated in the following way
\begin{equation}
 \begin{array}{c}
 \displaystyle
 \lim_{\zeta\to \ 0}\int\limits_0^\infty dxx^{2m\pm 1}e^{-(\zeta-2i)x}=
 \mp (-1)^m\frac{(2m\pm 1)!}{2^{2m+1\pm 1}}\ ,
 \\[3mm]
  \displaystyle
 2i\lim_{\zeta\to \ 0}\int\limits_0^\infty dxx^{2m}e^{-(\zeta-2i)x}=
  (-1)^{m+1}\frac{(2m)!}{2^{2m}}\ .
\end{array}
\label{19}
 \end{equation}
The physical interpretation of (\ref{19}) is as follows. In any
instant of time, the procedure of redefining the integrals in
(\ref{19}) selects the contribution of virtual gravitons with wave
lengths that are near characteristic value (\ref{9}) only, and
eliminates contributions of all other gravitons. This redefining
procedure provides the existence of the exact solution (\ref{8})
of BBGKY hierarchy.

Zero moment $W_0$ is not included in the equations of the theory.
Therefore, to avoid a logarithmic infrared singularity, the
function under the integral sign in (\ref{16}) is differentiated
initially, then derivatives are combined as $D=\ddot W_0+3H\dot
W_0$, and redefined integrals are calculated. The results are
\begin{equation}
 \begin{array}{c}
 \displaystyle
 D=-\frac{12\varkappa\hbar N_g}{\pi^2}H^4\ ,
 \qquad
 W_m=\frac{(-1)^{m+1}}{2^{2m}}
 (2m-1)!(2m+1)(m+2)\times
  \frac{2\varkappa\hbar N_g}{\pi^2}H^{2m+2} ,\qquad m\geqslant 1.
 \end{array}
 \label{20}
 \end{equation}
It is easy to verify that expressions (\ref{20}) satisfy to
recurrence relations (\ref{8}), following from the BBGKY hierarchy
immediately.

In accordance with (\ref{2}) and (\ref{20}), the energy density and the pressure of
virtual gravitons (that generate the back reaction of quantum
fluctuations on background geometry) are
\begin{equation}
\displaystyle \varepsilon_g=\frac{1}{12\varkappa}W_1=\frac{3\hbar
 N_g}{8\pi^2}H^4, \qquad p_g=-\varepsilon_g\ .
 \label{21}
 \end{equation}
One can also obtain the equation (\ref{21}) by other methods. The
vacuum amplitude of graviton field in the one--loop approximation
can be presented via a path integral, which, as it is known
\cite{11}, can be calculated exactly in the de Sitter space. One
can calculate this path integral under boundary conditions that
correspond to the state vector (\ref{13}), (\ref{15}), under the
assumption that the de Sitter geometry is self--consistent. A
variation of result of this calculation over the scale factor
leads to the convolution of Einstein equations (\ref{1}) and thus
to the expression $(\varepsilon_g -3p_g)/4\equiv \varepsilon_g $
that coincides with (\ref{21}). The same result can be obtained by
Schwinger--DeWitt method using the effective action formalism.

\section{Energy balance}

Equation (\ref{21}) is a consequence of the balance equation for
the energy that is emerging to the space due to graviton creation
and vanishing due to graviton annihilation. The characteristic
energy of gravitons in these processes is $\hbar \omega \sim \hbar
H$. Total probabilities of graviton creation and annihilation
processes (normalized to the unit volume) $w_{cr}$ and $w_{ann}$
are proportional to the phase volume of one graviton
$\omega^3/3\pi^2\sim H^3/3\pi^2$. The exponent of the
background--graviton coupling constant is unity if $\omega\sim H$.
Thus, we obtain for $w_{cr}$ and $w_{ann}$ the following
estimations
\begin{equation}
\begin{array}{c}
\displaystyle w_{cr}=\frac{\alpha}{3\pi^2}H^3(\overline{N}_{\bf
k}+1)(\overline{N}_{\bf -k}+1), \qquad
w_{ann}=\frac{\alpha}{3\pi^2}H^3\overline{N}_{\bf
k}\overline{N}_{\bf -k}.
 \end{array}
 \label{22}
 \end{equation}
where $\alpha=\mathcal{O}(1)$, $\overline{N}_{\pm{\bf k}}\sim
N_g/2$ is the average number of gravitons with wave lengths that
are near characteristic value (\ref{9}). Finally, we get the
balance equation in the form
 \begin{equation}
 \begin{array}{c}
\displaystyle \varepsilon_g =\hbar \omega
\left(w_{cr}-w_{ann}\right)= \frac{\alpha}{3\pi^2} \hbar H^4
(\overline{N}_{\bf k}+\overline{N}_{\bf -k}+1)
 \simeq  \frac{\alpha}{3\pi^2}\hbar N_g H^4.
 \end{array}
 \label{23}
  \end{equation}
This estimate with accuracy of numerical factor of the order of
unity coincides with (\ref{21}). From (\ref{22}) and (\ref{23})
one can see that the non-zero effect occurs because of quantum
spontaneous process of the particle creation.

If non-gravitational contributions to the vacuum energy are
compensated exactly (due to, e.g., a consequence of
Supersymmetry), it follows from (\ref{2}) and (\ref{21})
\begin{equation}
\displaystyle
 H^2=\frac{8\pi^2}{\varkappa\hbar N_g}\ .
 \label{24}
\end{equation}
According (\ref{24}), $H^2\sim \hbar^{-1}$, and it clearly
demonstrates that the acceleration of the Universe expansion is a
macroscopic quantum effect in the graviton vacuum at the scale of
the Universe (analogous to such known macroscopic quantum effects
as super-fluidity and super-conductivity). Substituting the
observable value of Hubble constant into (\ref{24}), we get an
estimate of the total number of virtual gravitons under the de
Sitter horizon $N_g\sim 10^{123}$.  This could be a possible
interpretation of the $\varepsilon_{{\scriptscriptstyle
D}{\scriptscriptstyle E}}/\varepsilon_{Plank}\sim 10^{-123}$
ratio.

\section{ Conclusion}

Virtual gravitons with wavelength of the order of the horizon must
appear and disappear in the graviton vacuum because of massless
and conformal non--invariance of the graviton field. A non-zero
balance of energy is due to the pure quantum process of
spontaneous graviton creation, in other words, due to the
uncertainty relation. Numerical integration of the system of
equations (\ref{2}), (\ref{6}) shows several types of solutions
with acceleration. In this paper, we considered the physics of
this effect by the example of the exact de Sitter solution, which
is the only one that corresponds to the state of equilibrium of
graviton vacuum, and this is the reason why this exact solution
exists.

{\bf Acknowledgment.} We would like to express our deep
appreciation to our friend and colleague Dr. Walter Sadowski for
invaluable advice and help in the preparation of the manuscript.

\end{document}